\documentclass{article}      
\usepackage{graphicx}
\begin{document}             
\title{Solving the Nos\'{e}-Hoover thermostat for Nuclear Pasta}
\author{M. \' Angeles P\'erez Garc\'ia \footnote{mperezga@usal.es}
\\
{\small { Departamento
de F\'{\i}sica Fundamental, Universidad de Salamanca}}
\\
{\small { Plaza de la Merced s/n 37008 Salamanca}, Spain } }

\date{January 21, 2006}      
\maketitle                   
\begin{abstract}
In this work we present a calculation of the hamiltonian variables 
solving the  molecular dynamics equations of motion
for a system of nuclear matter kept at fixed temperature 
by using the Nos\'{e}-Hoover Thermostat and interacting via a semiclassical 
potential depending on both positions and momenta. 
\end{abstract}
\section{Introduction}

At densities just below nuclear saturation density, there may be spatial
structures different from the uniformly distributed matter.
In events like a supernova core collapse large quantities of neutrinos are produced.
As they stream out of the star, coherent scattering out of spatial density 
fluctuations of neutron rich matter, known as {\it pasta phases},
 can happen. These structures arise due to
the competition among intermediate-range attractive and long-range repulsive forces 
\cite{pasta}. Possible shapes include round nuclei, flat plates, rods, and spherical voids
\cite{pasta_shapes}.

 In this work we are interested in solving the dynamical equations for a nuclear
 system interacting via a realistic potential model depending not only on positions,
 as usual, but on both positions and momenta and keeping the temperature fixed using
the Nos\'{e}-Hoover thermostat\cite{nh}. We mimic the Pauli principle in a fermionic
nucler system by adding a momentum dependent term to the potential \cite{dorso}. In this way the 
characteristic phase
space repulsion for fermionic nucleons (protons and neutrons) can be included, restricting some
of the available dynamical states for the individual particles. 
 Solving this system allows for realistic configurations of neutron rich plasmas in the 
NVT ensemble.


\section{Model Hamiltonian and Method}

We model a charge-neutral system with a fixed number of nucleons,
 $A$, and electrons.  The electrons provide a neutralizing background and 
are described as an almost degenerate free Fermi gas, see below.
The Hamiltonian under the Nos\'e-Hoover method
for the extended system in this case can be written as \cite {nh}
\begin{eqnarray}
  { H}_{\rm NH} &=& \sum_{i=1}^{A} \frac{{\bf P}_{i}^{2}}{2m_{i}}
  + {V}({R}_{ij}, {P}_{ij})
  + \frac{s^2 p_{s}^{2}}{2Q} + g \frac{\ln s}{\beta}\,
\label{hnh}
\end{eqnarray}
where ${V}({R}_{ij}, {P}_{ij})$ is the potential
which depends on both positions and momenta, described below,
$s$ is the extended position variable, 
$p_{s}$ is the momentum conjugate to $s$,
$Q$ is the thermal inertial parameter
corresponding to a coupling constant between the system and the thermostat
taking a value  $Q\sim 10^6$-$10^8$ MeV $(fm/c)^2$,
we use $g=3A$ as a condition for generating the canonical ensemble
in the classical molecular dynamics simulations, $\xi$ is
the thermodynamic friction coefficient
and $\beta$ is defined as $\beta= 1/k_{\rm B}T$.

The total potential energy of the system, $V$, consists of a
sum of two-body interactions
\begin{equation}
 V= V_{had}+ V_{Coulomb} + V_{Pauli}  \
\label{vtot}
\end{equation}
where
\begin{equation}
V_{had} =\sum_{i<j} a e^{-R_{ij}^{2}/\Lambda} + \Big[b+c\tau_{i}\tau_{j}\Big]
       e^{-R_{ij}^{2}/2\Lambda}\
\label{vhad}
\end{equation}
\
\begin{equation}  
V_{Coulomb}=\sum_{i<j} \frac{e^{2}}{R_{ij}}e^{-R_{ij}/\lambda}
                 \frac{(1+ \tau_{i})}{2} \frac{(1+ \tau_{j})}{2} \;,
\label{vc}
\end{equation}

\begin{equation}  
 V_{Pauli}=
  d \left( \frac{\hbar}{q_0 p_0} \right )^3
  \sum_{i, j(\neq i)}
  \exp{ \left [ -\frac{({R}_{ij})^2}{2q_0^2}
          -\frac{({P}_{ij})^2}{2p_0^2} \right ] }\
  \delta_{\tau_i \tau_j} \delta_{\sigma_i \sigma_j}\ ,
\label{vpauli}
\end{equation}

Here the distance between the particles in phase space is denoted by $R_{ij}=|{\bf
R}_i\!-\!{\bf R}_j|$, $P_{ij}=|{\bf P}_i\!-\!{\bf P}_j|$  and $\tau_{i}$ represents 
the ith-nucleon isospin 
projection on z-axis ($\tau=\!+\!1$ for protons and $\tau=\!-\!1$ for
neutrons). 
$V_{Coulomb}$ corresponds to the screened Coulomb interaction.
The screening length, $\lambda$, that results from the slight polarization of the
electron gas is arbitrarily set to $\lambda\!=\!10~fm$ as in previous works 
\cite{pasta1}\cite{pasta2} \cite{pasta3}.
$V_{Pauli}$ 
is the Pauli potential that incorporates phase space repulsion for fermions by means 
of the  Kronecker deltas in spin and isospin \cite{peilert}.
This interaction model contains the characteristic
intermediate-range attraction and short-range repulsion of the
nucleon-nucleon force through $V_{had}$ and allows to include the fermionic 
nature of nucleons. 

The parameter set employed is displayed in table~\ref{parset}, adjusted
to reproduce the saturation density and binding
energy per nucleon of symmetric nuclear matter and neutron matter, 
and the binding energy of finite nuclei at $T=1$ MeV.

\begin{table}[h]
\caption{Model parameters.}
 \begin{tabular}{ccccccc}
  $a$ (MeV) & $b$ (MeV)& $c$(MeV) & 
  $d$(MeV) & $q_{0}$(fm) & $p_{0}$(MeV/c) & $\Lambda$ (fm$^{2}$) \\
  \hline
  133 & -47 & 11 & 29 & 3 & 120 & 1.5 
 \label{parset}
 \end{tabular}
\end{table}

According to the hamiltonian eq.(\ref{hnh}) the equations of motion for each nucleon yield
\begin{eqnarray}
  \frac{d {\bf R}_{i}}{dt}
  &=& \frac{\partial{ H}_{\rm NH}}{\partial {\bf P}_{i}}
  = \frac{{\bf P}_{i}}{m_{i}}
  + \frac{\partial{V}}{\partial {\bf P}_{i}}\ ,\\
  \nonumber\\
  \frac{d {\bf P}_{i}}{dt}
  &=& -\frac{\partial{ H}_{\rm NH}}{\partial {\bf R}_{i}}
  = -\frac{\partial{V}}{\partial{\bf R}_{i}} - \xi {\bf P}_{i}\ ,\\
  \nonumber\\
  \frac{1}{s} \frac{ds}{dt}
  &=& \frac{1}{s} \frac{\partial{ H}_{\rm NH}}{\partial p_{s}}
  = \frac{1}{Q} \frac{\partial{ H}_{\rm NH}}{\partial\xi} = \xi\ ,\\
  \nonumber\\
  \frac{d\xi}{dt}
  &=& \frac{1}{Q} \left\{ \sum_{i=1}^{A}\left(
      \frac{{\bf P}_{i}^{2}}{m_{i}}
      + {\bf P}_{i}\cdot\frac{\partial{V}}{\partial{\bf P}_{i}}
      \right) - \frac{g}{\beta} \right\},\label{dxi}
\end{eqnarray}

with $\xi \equiv \frac{sp_{s}}{Q}\ $

We then study the solution of the  equations of motion using different methods,
 like the exponential fitting \cite{vigo2}, 
exponentially fitted symplectic method \cite{vigo3},
  but we choose the Numerov type
integrator algorithm \cite{vigo} \cite{vigo1} because it allows 
the best accuracy, giving order 8 or 12 for this kind of hamiltonians.

The energy of the thermostat system is a conserved quantity
\begin{equation}
  E_{NH}= \sum_{i=1}^{A} \frac{ P_{i}^2}{2m} +V+\frac{1}{2 }Q{\xi}^2+gkTlns=
K+V+E_{\xi}+E_T
  \label{e_ext}
\end{equation}

An effective temperature, that fluctuates around the desired initially set temperature
 $T$, can be defined as
\begin{equation}
  T_{\rm eff}=
  \frac{2}{3A k_{\rm B}}  \sum_{i=1}^{A} \frac{1}{2}{\bf P}_{i} \cdot
  \frac{d {\bf R}_{i}}{dt}\
  \label{teff}
\end{equation}

\vspace{0.2in}
\begin{figure}[hbtp]
\begin{center}
\includegraphics [scale=0.5] {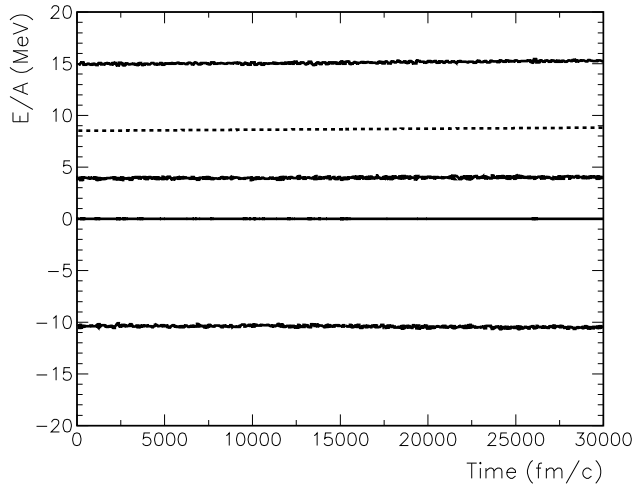}
\caption{ Energy per nucleon versus simulation time at $T=1~MeV$, 
$n_{b}=0.016 fm^{-3}$ and $Y_e=0.2$. From top to bottom we plot
 $E_T,E_{NH},K,E_{\xi},V$ per nucleon.}
\label{Fig1}
\end{center}
\end{figure}

\section{Simulation Results}
\label{results}

The simulations presented in this work were carried out with fixed baryonic 
number density, $n_{b}$, 
lepton fraction, $Y_{e}=\frac {n_{e}}{n_b}$  and 
a fixed number of nucleons, $A$, initially placed in a cubic box of side
$L=(A/n_{b})^{1/3}$ at random. To minimize finite size effects we use periodic boundary conditions. 
Up to 1,000 particles are taken in this work with typical thermalization times of order $10^5 fm/c$. 

As an example of typical low density conditions we consider
$n_{b}=0.016 fm^{-3}$, which is about a tenth of normal nuclear
density, a temperature $T=1~MeV$ and a typical electron fraction for neutron rich matter
 $Y_e=0.2$.

The energy of the system is conserved, according to  eq.~(\ref{e_ext}), as can be seen in
the plot of energy per particle versus time in Fig.~\ref{Fig1}.
The system exhibits characteristic oscillations in the thermostat variables due 
to the value of $Q$, that it is associated with the heat capacity of the system. 

\vspace{0.2in}
\begin{figure}[hbtp]
\begin{center}
\includegraphics [scale=0.5] {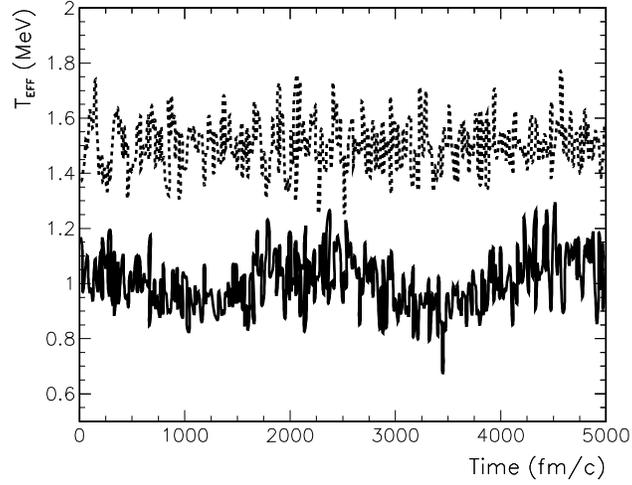}
\caption{ Effective temperature for a configuration of 200 particles at $T_{eff}=1~MeV$, baryon density
$n_{b}=0.016 fm^{-3}$ and $Y_e=0.2$ for $Q=10^6~MeV(fm/c)^2$ and $Q=10^8~MeV 
(fm/c)^2$. See text for details.}
\label{Fig2}
\end{center}
\end{figure}

In Fig.~\ref{Fig2} we plot the effective
 temperature for a system of $A=200$ particles at $n_{b}=0.016 fm^{-3}$, 
$Y_e=0.2$ and two values of Q. The upper curve (dashed line) corresponds 
to 
$Q=10^6$ MeV $(fm/c)^2$ and the lower curve (solid line) to $Q=10^8$ MeV $(fm/c)^2$. The 
temperature is set to T=1 MeV for both curves. The upper curve has been biased by adding an
 offset of $0.5$ MeV for the sake of clarity. 
By decreasing Q the temperature control is better but leads to rapid oscillations in the
energy, which must be carefully considered when studying the dynamical response in energy 
modes of the system \cite{dyn}.

\begin{figure}[hbtp]
\begin{center}
\includegraphics [scale=0.5] {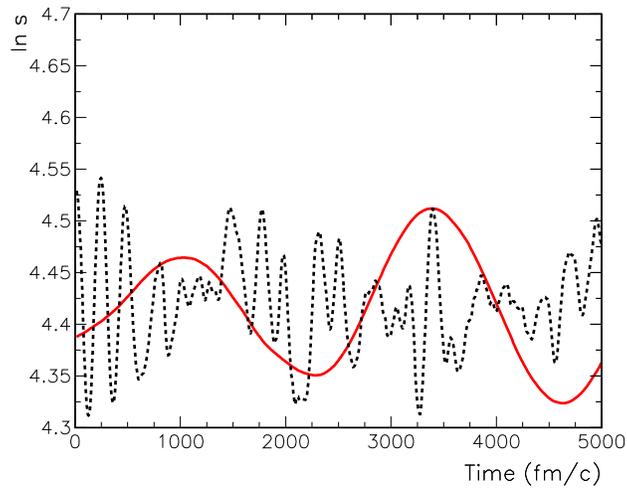}
\caption{ Thermostat variable $ln s$ versus simulation time at $T_{eff}=1~MeV$, baryon density
$n_{b}=0.016 fm^{-3}$ and $Y_e=0.2$ for $Q=10^6~MeV(fm/c)^2$ and $Q=10^8~MeV 
(fm/c)^2$. See text for details. }
\label{Fig3}
\end{center}
\end{figure}

In the same way in Fig.~\ref{Fig3} we plot the thermostat variable ln s for the same cases as
in Fig.~\ref{Fig2} and
again the dashed line corresponds to 
$Q=10^6$ MeV $(fm/c)^2$ and the solid line to $Q=10^8$ MeV $(fm/c)^2$. This illustrates 
the slow time variation of $E_T$ for systems where Q is set bigger.

\vspace{0.2in}

In the numerical simulation the energy of the system may suffer a small drift with time
due to the accuracy of the algorithm used. 
We have checked that the maximum relative extended energy error at time t, defined as
\begin{equation}
\frac{\Delta E}{E}= \left |\frac{E(t)-E(0)}{E(0)}\right |
\end{equation}
is $\frac{\Delta E}{E}=10^{-4}$ for time length $\Delta t=5.10^3~fm/c$ using timesteps $dt=0.025~fm/c$
and  $Q=10^6~MeV(fm/c)^2$. This error increases somewhat with timestep size. 
We use timesteps in the range $dt=0.01-0.1~fm/c$ for our simulations in this work.

\section{Conclusions}
\label{conclusions} 
In this work we have employed molecular dynamics techniques to solve 
a hamiltonian model with an interaction potential 
depending on both positions and momenta.  We have simulated
neutron rich matter at a given density and kept fixed the temperature by using the
Nos\'{e}-Hoover thermostat and solved the equations of motion using different integration
methods. We find that a Numerov type algorithm allows 
the best accuracy for this kind of hamiltonians giving order 8 or 12.
  
We find that at the density of a tenth of nuclear saturation
density, $Y_e=0.2$ and T=1 MeV a clustered phase is formed.
By changing the heat capacity of the system in the range $Q=10^6-10^8~MeV(fm/c)^2$ 
thermalization of the system is achieved in times of order  $10^5~fm/c$.
The lower the heat capacity, Q, the better the temperature control is.
Induced oscillations in the thermostat variables must
be considered with further detail as they could jeopardize low energy excitation modes.

\section*{Acknowledgments}
The author wishes to thank Charles J. Horowitz, Jorge Piekarewicz, J. Vigo-Aguiar and B. Wade 
for helpful comments. This work has been partially supported by the Spanish Ministry of
education  under project BFM2003-021121.

%


\begin{thebibliography}{99}

\bibitem{pasta} D. G. Ravenhall, C. J. Pethick, and J. R. Wilson,
                 Phys. Rev. Lett. {\bf 50}, 2066 (1983).
                 M. Hashimoto,
                H. Seki, and M. Yamada, Prog. Theor. Phys.
                {\bf 71}, 320 (1984).

\bibitem{pasta_shapes}Gentaro Watanabe,
                       Katsuhiko Sato, Kenji Yasuoka, and
                       Toshikazu Ebisuzaki,
                       Phys. Rev. C~{\bf 69} 055805 (2004)

\bibitem{nh}  S. Bond, B. Leimkulher, B. Laird, Journal of Computational
              Physics ~{\bf 151} 114 (1999) and references therein.

\bibitem{dorso}  C. Dorso, S. Duarte, J. Randrup, Phys Lett B~{\bf 215}(1988) 611 

\bibitem{pasta1} C. J. Horowitz, M. A. Perez-Garcia, and J. Piekarewicz,
                 Phys. Rev. C~{\bf 69}, 045804 (2004).

\bibitem{pasta2} C. J. Horowitz, M. A. Perez-Garcia, J. Carriere,
                 D. K. Berry, and J. Piekarewicz,
                 Phys. Rev. C~{\bf 70} 065806 (2004).

\bibitem{pasta3} C. J. Horowitz, M. A. Perez-Garcia, 
                 D. K. Berry, and J. Piekarewicz,
                 Phys. Rev. C~{\bf 72} 035801 (2005).
\bibitem{peilert} G. Peilert, J. Randrup, H. Stocker, W. Greiner
                 Phys. Let. B~{\bf 260}, 271 (1991).

\bibitem{vigo2} J. Vigo-Aguiar, J. Ferrandiz,  SIAM J Numer Anal 35 (4), 1684 (1998) 
\bibitem{vigo3} Simos T.E., J. Vigo-Aguiar, J Phys Rev  E 67 (1) 016701 (2003 )
\bibitem{vigo}  J. Vigo-Aguiar, H. Ramos, Math. and Computer Modelling, Vol42, 7, 837, (2005)
\bibitem{vigo1}  J. Vigo-Aguiar, H. Ramos, Journal of Math. Chemistry, 37, 3, 252, (2005)

\bibitem{dyn} C. J. Horowitz, M. A. Perez-Garcia, preprint.

\end{thebibliography}
\end{document}